# The electroweak phase transition on the lattice

Zoltán Fodor*

Deutsches Elektronen-Synchrotron, DESY, 22603 Hamburg, Germany

**Abstract**

The finite temperature electroweak phase transition is studied on the lattice. The results of the simulations obtained by the 3-dimensional effective theories and the 4-dimensional SU(2)-Higgs model are reviewed.

## 1. Introduction

The perturbative sector of the electroweak (EW) theory is extremely successful and we have even reached a point at which the Higgs mass ($m_H$) starts to appear in the EW precision data. However, there are basic questions in the theory, which can not be answered within the perturbative framework. One of them is the finite temperature electroweak phase transition.

At high temperatures ($T$) the spontaneously broken EW symmetry is restored. Since the baryon-number violating processes are unsuppressed at high $T$, the cosmological EW phase transition plays a crucial role in the understanding of the observed baryon asymmetry [1]. The idea of the EW baryogenesis needs a departure from thermal equilibrium, thus a sufficiently strong first order phase transition via bubble nucleation. A strong first order phase transition could explain the observed asymmetry, a weak one could have washed out any B+L asymmetry.

Unfortunately, the perturbative treatment of the phase transition suffers from infrared problems. In the realistic Higgs mass range ($m_H > 63\ GeV$) the perturbative approach predicts $\mathcal{O}(100\%)$ corrections [2, 3, 4] for the relevant quantities (e.g. interface tension or latent heat). Some of the nonperturbative estimates (e.g. magnetic mass [2]) suggest a weaker, others (e.g. vacuum-condensate [5]) a stronger first order phase transition than the perturbative approach. The only way to solve the problem seems to be the use of Monte-Carlo simulations on the lattice.

Since fermions always have nonzero Matsubara frequencies, the perturbative treatment of these, at high temperatures very massive, modes could be satisfactory. Therefore, the starting point of the lattice analyses is the $SU(2)$-Higgs model, which contains all the essential features of the standard model of electroweak interactions.

In Sect. 2 the basic idea of the dimensional reduction and the different reduced 3-dimensional models are presented. Sect. 3 deals with the results of the simulations (3-dimensional ferromagnet model [6], 3-dimensional gauge-Higgs model [7], 4-dimensional, finite temperature $SU(2)$-Higgs model [8, 9]).

* On leave from Institute for Theoretical Physics, Eötvös University, Budapest, Hungary

## 2. 4-dimensional $SU(2)$-Higgs model and effective 3-dimensional models

The 4-dimensional $SU(2)$-Higgs model at finite $T$ is defined by the following action

$$S = \int_0^\beta d\tau \int d^3x \left[\frac{1}{4}F^a_{\mu\nu}F^a_{\mu\nu} + (D_\mu\phi)^\dagger(D_\mu\phi) - \frac{1}{2}m^2\phi^\dagger\phi + \lambda(\phi^\dagger\phi)^2\right], \qquad (1)$$

where $D_\mu$ and $F^a_{\mu\nu}$ are the usual covariant derivative and the Yang-Mills field strength, respectively. $\beta = 1/T$, and the $\tau$ integration is over periodic bosonic fields. This model has been studied on the lattice by ref [8, 9].

The origin of the perturbative infrared problems is the appearance of zero Matsubara modes in the bosonic sector. Therefore, similarly to the case of the fermion fields with nonzero Matsubara frequencies one can integrate out all the massive, non-static bosonic modes of (1) at the one-loop level. Since these modes are heavy in the high temperature limit, the perturbative treatment of them could be well-founded. This dimensional reduction gives an effective 3-dimensional gauge-Higgs model, where in addition, also an isovector field ($A_0^a$), the fourth component of the gauge fields, is present

$$S = \int d^3x \left[ \frac{1}{4} F_{ij}^a F_{ij}^a + \frac{1}{2} (D_i A_0)^a (D_i A_0)^a + (D_i\phi)^\dagger (D_i\phi) + \frac{1}{2} m_D^2 A_0^a A_0^a + \frac{1}{4}\lambda_A (A_0^a A_0^a)^2 + m_3^2 \phi^\dagger\phi + \lambda_3 (\phi^\dagger\phi)^2 + \frac{1}{2} h_3 A_0^a A_0^a \phi^\dagger \phi \right]. \quad (2)$$

Here all the 3-dimensional couplings $g_3^2$, $\lambda_3$, $\lambda_A$ and $h_3$ have dimension $[GeV]$. These parameters and the masses $m_D$ and $m_3$ can be expressed in terms of the 4-dimensional couplings and the temperature [10, 11]. This model has been studied on the lattice by ref [7].

The next step is the elimination of the gauge degrees of freedom. One obtains a a 3-dimensional $\mathcal{O}(4)$ ferromagnet model with cubic and quartic terms in the action.

$$S = \int d^3x \left[ (D_i\phi)^\dagger (D_i\phi) + \bar{m}_3^2 \phi^\dagger\phi + \bar{\lambda}_3 (\phi^\dagger\phi)^2 - 2\bar{g}_3^2 (\bar{m}_T^2 + \phi^\dagger\phi)^{3/2} - \bar{g}_3^2 (\bar{m}_E^2 + \phi^\dagger\phi)^{3/2} \right]. \quad (3)$$

The parameters of this action can be similarly expressed in terms of the 4-dimensional couplings and the temperature as in the previous case [6]. The infrared stability is ensured by including magnetic ($\bar{m}_T$) and electric ($\bar{m}_E$) screening mass terms. Since the first of them has a nonperturbative origin, the approximation contains an uncertainty connected with the magnetic mass. This model has been studied on the lattice by Ref. [6]. Their choice of $\bar{m}_T$ has been suggested by the solution of the one-loop gap-equations of the $SU(2)$-Higgs theory at finite temperature [2].

## 3. Lattice simulations

The results obtained by the different groups are summarized. First the results of the 3-dimensional $\mathcal{O}(4)$ ferromagnet model, then those of the 3-dimensional gauge-Higgs model are presented. Finally the analyses based on the 4-dimensional $SU(2)$-Higgs model are considered.

### *3.1. The 3-dimensional $\mathcal{O}(4)$ ferromagnet model*

After discretizing the action (3) and performing the mean field analysis of this effective scalar model Ref. [6] presented the results of their Monte-Carlo simulations. The typical lattice sizes and statistics for a given lattice size were $8^3 - 18^3$ and $10^6$ sweeps, respectively. The parameters used in this work correspond to the physical values of the $W$ mass and vacuum expectation value of the Higgs field ($m_W = 80\ GeV$ and $v = 246\ GeV$), however, the Higgs mass ($m_H \approx 35\ GeV$) was considerably smaller than the experimental bound. This value has been selected: i) to have a strong signal of a first order transition; ii) to be free of the uncertainties due to renormalisation prescriptions; iii) to be able to compare the results with other 3-dimensional analyses, e.g. [7].

The simulations have shown a considerably weaker first order phase transition than the mean-field analysis. For the critical temperature ($T_c$), jump in the order parameter at the critical temperature ($\varphi_c$), latent heat ($\Delta\epsilon$) and interface tension ($\sigma$) the simulations have given $T_c = 114.6(36)$, $\varphi_c/T_c = 0.68(4)$, $\Delta\epsilon/T_c^4 = 0.122(8)$ and $\sigma/T_c^3 \approx 6.4 \cdot 10^{-4}$, respectively. The corresponding mean field results are $T_c = 99.6$, $\varphi_c/T_c = 1.3$, $\Delta\epsilon/T_c^4 = 0.262$ and $\sigma/T_c^3 = 0.024$, respectively.

Since the non-perturbative dynamics of the scalar fields seems to weaken the phase transition, any observation of hard first order phase transition could result only from the non-perturbative effects related to the gauge degrees of freedom.

### *3.2. The 3-dimensional gauge-Higgs model*

In [11] the one- and two-loop effective potential is constructed using the dimensional reduction idea (cf. eq. 2). With the help of the renormalisation group leading logarithms have been summed.

The lattice results have been presented in [7]. The used $W$ mass and gauge coupling were $m_W = 80.6\ GeV$ and $g = 2/3$, respectively. The runs have been done on lattices of sizes $8^3 - 32^3$.

In the broken phase the perturbative and lattice results are in very good agreement, e.g. the predictions for the expectation value of the Higgs field agree within 1%. However, for other quantities, which are crucial for the cosmological phase transition, the lattice simulations suggest a stronger first order phase transition than the perturbative approach. For $m_H = 35\ GeV$ the critical temperatures obtained on the lattice and in the perturbative approach are $T_c = 85\ GeV$ and $T_c = 95\ GeV$, respectively; for $m_H = 80\ GeV$ the values are $T_c = 162.1(26)\ GeV$ and $T_c = 173.3\ GeV$, respectively. The jump in the order parameter is $\varphi_c/T_c = 0.73(4)$ obtained by the lattice simulation and $\varphi_c/T_c = 0.47$ in perturbation theory.

The informations obtained from the above lattice studies are in qualitative agreement with the vacuum-condensate picture [5]. Using the vacuum energy-shift suggested by the above data and following [5], the electroweak baryogenesis could be possible up to Higgs mass of about 80 $GeV$. The exact determination of this bound, however, needs further study.

### *3.3. 4-dimensional $SU(2)$-Higgs model at finite temperature I.*

The most straightforward way to study the problem on the lattice is discretizing the action of (1) and performing the simulations on asymmetric thermal lattices, $L_t \ll L_x, L_y, L_z$ ($L_t$ is the extension of the lattice in the temperature direction).

In [8] the typical lattice sizes were $2 \cdot 16^3 - 2 \cdot 36^3$ and approximately 7000 measurements have been done for each set. The simulations have been performed for $m_W = 80\ GeV$, $m_H \approx m_W$ and $g^2 = 0.5$. The lattice results (two-state signal, multihistogram and finite size analysis) show a clear first order phase transition. The phase transition is stronger than the one obtained in the one-loop perturbative approach. The jump in the order parameter is $\varphi_c/T_c = 0.68$ given by the lattice simulation and $\varphi_c/T_c = 0.3$ in perturbation theory. The latent heat and the metastability temperature range ($\delta T/T_c$) exceeds perturbative estimates by an order of magnitude, thus e.g. $\delta T/T_c = 0.076$ on the lattice and $\delta T/T_c = 0.008$ in the perturbative approach.

### *3.4. 4-dimensional $SU(2)$-Higgs model at finite temperature II.*

A 4-dimensional, finite $T$ analysis has ben done by [9] too. The largest lattice used was $2 \cdot 32^2 \cdot 256$ with $\mathcal{O}(10^5)$ updating. The simulations have been performed for two set of parameters: $m_W = 80\ GeV$, $m_H = 18\ GeV, g^2_{bare} = 0.5$ (*low*) point; $m_W = 80\ GeV$, $m_H = 49\ GeV, g^2_{bare} = 0.5$ (*high*) point. A moderate renormalisation of the gauge-coupling, $g_R^{2,low} = 0.5476(90)$ and $g_R^{2,low} = 0.5781(95)$, has been found.

The latent heat has been calculated from the discontinuity of the energy density ($\Delta\epsilon$). The necessary partial derivatives along the *lines of constant physics* have been determined by use of the one-loop renormalisation group equations, whereas the critical points have been given by the inspection of the gauge-invariant effective potential [12, 13, 4]. The obtained values are: $(\Delta\epsilon/T_c^4)^{low} = 1.68(17)$ and $(\Delta\epsilon/T_c^4)^{high} = 0.125(19)$.

The interface tension has been determined using the two-coupling method of Potvin and Rebbi [14]. The obtained results are: $(\sigma/T_c^3)^{low} = 0.84(16)$ and $(\sigma/T_c^3)^{high} = 0.008(2)$.

The lattice results for $\Delta\epsilon$ and $\sigma$ are in qualitative good agreement with the two-loop resummed perturba-tive results [3] (see Fig. 1). Note, that the $L_t = 3$ simulations [15] confirm the above conclusions.

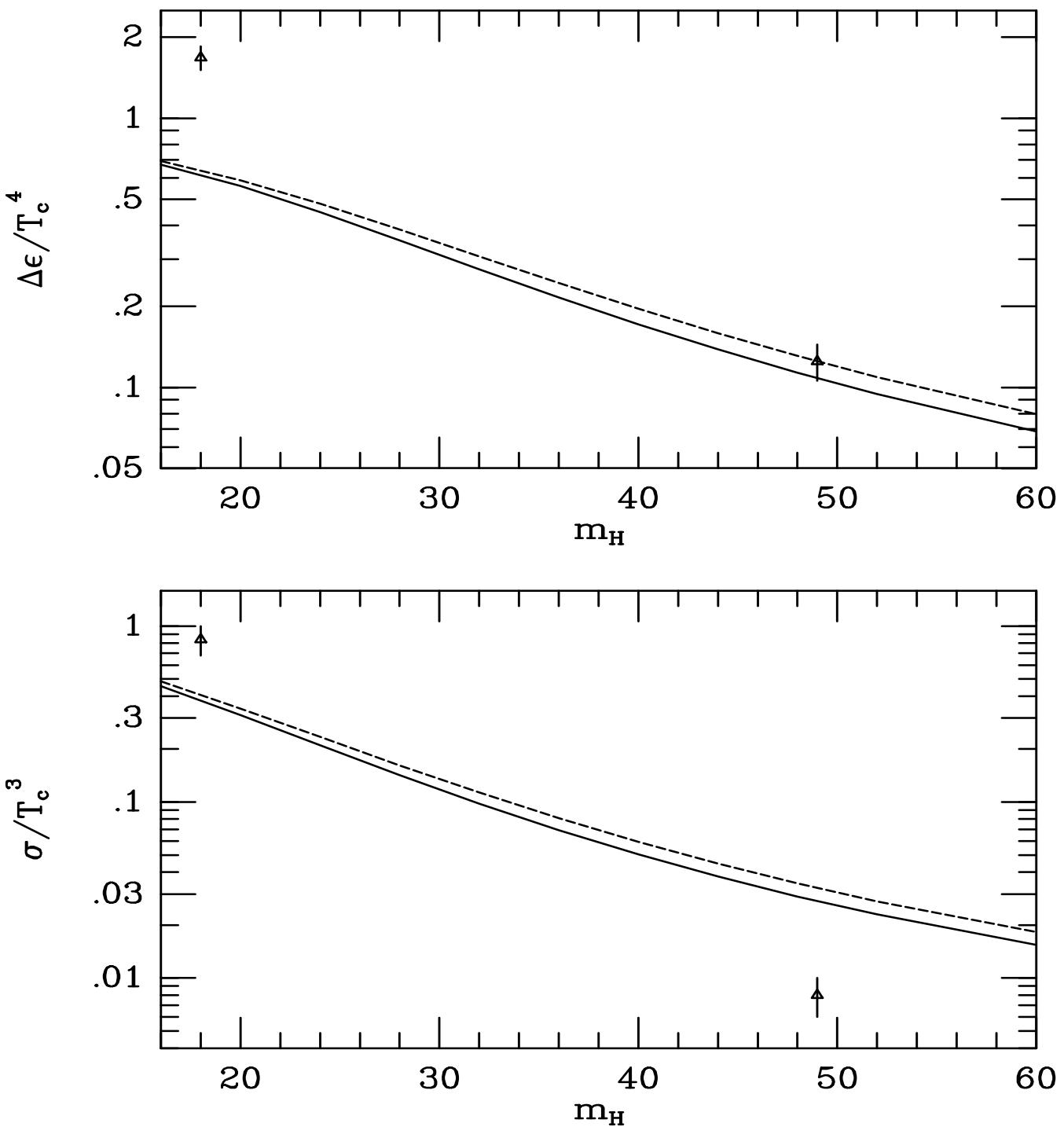

**Figure 1.** The perturbative and the lattice results for $\Delta\epsilon$ and $\sigma$ for different Higgs masses. The solid line corresponds to $g_R^{low}$ the dashed one to $g_R^{high}$.

## References

[1] V. A. Kuzmin, V. A. Rubakov and M. E. Shaposhnikov, Phys. Lett. **B155** (1985) 36.

[2] W. Buchmüller, Z. Fodor, T. Helbig and D. Walliser, Ann. Phys. (NY) **234** (1994) 260.

[3] Z. Fodor, A. Hebecker, DESY-94-025 (1994), Nucl. Phys. B, in press.

[4] Z. Fodor, these proceedings.

[5] M. Shaposhnikov, Phys. Lett. **B316** (1993) 112.

[6] F. Karsch, T. Neuhaus, A. Patkós BI-TP 94/27.

[7] K. Kajantie, K. Rummukainen, M.E. Shaposhnikov, Nucl. Phys. **B407** (1993) 356; K. Farakos, K. Kajantie, K. Rummukainen, M.E. Shaposhnikov, CERN-TH.7244/94 (1994).

[8] B. Bunk, E. M. Ilgenfritz, J. Kripfganz, A. Schiller, Nucl. Phys. **B403** (1993) 453.

[9] F. Csikor, Z. Fodor, J. Hein, K. Jansen, A. Jaster, I. Montvay, Phys. Lett. **B334** (1994) 405.

[10] A. Jakovác, K. Kajantie, A. Patkós, Phys. Rev. **D49** (1994) 6810;.

[11] K. Farakos, K. Kajantie, K. Rummukainen, M.E. Shaposhnikov, CERN-TH.6973/94 (1994).

[12] M. Lüscher, unpublished notes (1988).

[13] W. Buchmüller, Z. Fodor and A. Hebecker, Phys. Lett. **B331** 131 (1994).

[14] J. Potvin, C. Rebbi, Phys. Rev. Lett. **62** (1989) 3062.

[15] Z.Fodor, J. Hein, K. Jansen, A. Jaster, I. Montvay, DESY-94-159.